\documentclass[12pt]{iopart}

\usepackage{graphicx}
\usepackage{subfig}
\usepackage{comment}
\usepackage[countmax]{subfloat}

\newcommand{\ynp}{Yb$_2$Ni$_{12}$P$_7$}

\begin{document}

\title[Crossover between Fermi liquid and non-Fermi liquid behavior in Yb$_2$Ni$_{12}$P$_7$] {Crossover between Fermi liquid and non-Fermi liquid behavior in the non-centrosymmetric compound Yb$_2$Ni$_{12}$P$_7$}

\author{S Jang$^{1,2,3}$, B D White$^{1,2}$, P -C Ho$^{4}$, N Kanchanavatee$^{1,2}$,  M Janoschek$^{1,2}$\footnote{Current address: Los Alamos National Laboratory, Los Alamos, NM 87545, USA}, J J Hamlin$^{1,2}$\footnote{Current address: Department of Physics, University of Florida, Gainesville, FL 32611, USA} and M B Maple$^{1,2,3}$}
\address{$^1$ Department of Physics, University of California, San Diego, La Jolla, California 92093, USA}
\address{$^2$ Center for Advanced Nanoscience, University of California, San Diego, La Jolla, California 92093, USA}
\address{$^3$ Materials Science and Engineering Program, University of California, San Diego, La Jolla, California 92093, USA}
\address{$^4$ Department of Physics, California State University, Fresno, CA 93740-8031, USA}
\ead{mbmaple@ucsd.edu}

\begin{abstract}

A crossover from a non-Fermi liquid to a Fermi liquid phase in Yb$_2$Ni$_{12}$P$_7$ is observed by analyzing electrical resistivity $\rho(T)$, magnetic susceptibility $\chi(T)$, specific heat $C(T)$, and thermoelectric power $S(T)$ measurements.  The electronic contribution to specific heat, $C_{e}(T)$, behaves as $C_{e}(T)/T \sim -\ln(T)$ for 5 K $< T <$ 15 K, which is consistent with non-Fermi liquid behavior.  Below $T \sim$ 4 K, the upturn in $C_{e}(T)/T$ begins to saturate, suggesting that the system crosses over into a Fermi-liquid ground state.  This is consistent with robust $\rho(T) - \rho_0 = AT^2$ behavior below $T \sim$ 4 K, with the power-law exponent becoming sub-quadratic for $T >$ 4 K.  A crossover between Fermi-liquid and non-Fermi liquid behavior suggests that Yb$_2$Ni$_{12}$P$_7$ is in close proximity to a quantum critical point, in agreement with results from recent measurements of this compound under applied pressure.

\end{abstract}


\submitto{\JPCM}
\maketitle

\section{INTRODUCTION}
\label{sec:introduction}

\indent For more than four decades, rare-earth and actinide intermetallic compounds containing Ce, Pr, Sm, Eu, Tm, Yb, and U have received considerable attention due to their unusual properties, which are generally related to, or consequences of, an $f$-electron valence instability.  The low-temperature properties are associated with hybridization of the localized $f$-electron and conduction electron states\cite{Sereni91,Fisk88} as described by the Friedel-Anderson model.\cite{Anderson61}  There are many models for this subtle electronic state that are neither localized nor truly itinerant, but a common feature is the existence of a characteristic energy scale $k_{B}T^{*}$ that delineates high-temperature magnetic (Curie-Weiss law) behavior from low-temperature nonmagnetic (Pauli-like) behavior of the $f$-electrons.  When the strength of the hybridization is especially large, the $f$-electron shell may fluctuate between two $f$-electron configurations $f^{n}$ and $f^{n-1}$, where $n$ is an integer.  The unit-cell volume of a system with rare-earth ions in such a ``valence-fluctuating'' state\cite{Lee87} is intermediate between those of the integral-valence states that correspond to the configurations $f^{n}$ and $f^{n-1}$.  This results in a deviation of the unit-cell volume from the corresponding volumes of trivalent lanthanide ions associated with the lanthanide contraction.  Furthermore, the electrical resistivity, $\rho(T)$, of valence fluctuating systems typically shows a Fermi liquid-like quadratic temperature dependence at low temperature and a shoulder in $\rho(T)$ at high temperatures.\cite{Lee87,Buffat86,Fujii89,Takabatake90}

The element Yb has an ambivalent character in a number of metallic compounds, resulting in an intermediate-valence state consisting of a temporal admixture of Yb$^{3+}$ ($4f^{13}$) and Yb$^{2+}$ ($4f^{14}$).\cite{Gsheidner69}  In particular, Yb compounds are also of interest because the trivalent Yb ion can be thought of as the 4$f$ hole counterpart of the trivalent Ce ion, which has one electron in its 4$f$ shell.  As in the case of Ce compounds,\cite{Klasse81} the Yb-based intermetallic compounds exhibit a rich diversity of physical properties that are not completely understood.\cite{Sarrao99}

The low-temperature ground states of $f$-electron compounds, and the quantum phase transitions between them, often generate striking quantum phenomena and phases.  Quantum phase transitions, or quantum critical points (QCPs), and their effect on the physical properties at finite temperatures have attracted considerable attention since the 1990's.\cite{Stewart01}  In metallic systems, strong deviations from Fermi-liquid (FL) behavior in the neighborhood of a QCP have been observed and are manifested by pronounced non-Fermi-liquid (NFL) behavior.  For example, the electronic contribution to specific heat divided by temperature, $C_{e}/T$, shows characteristic $-\ln T$ behavior at low temperatures instead of becoming temperature independent.  The electrical resistivity, $\rho(T)$, is proportional to $T^n$ with an exponent of $n$ = 1 - 1.5 instead of the FL result where $n$ = 2.\cite{maple10}  While a large number of Ce- and U-based heavy-fermion (HF) systems have been discovered and thoroughly investigated,\cite{Stewart01} there are comparatively fewer Yb-based HF systems in which the low-temperature ground states have been studied.  Only a few of these appear to be located near a QCP, and in all of those cases, the QCP is associated with suppression of magnetic order (one prominent exception is $\beta$-YbAlB$_4$ in which a QCP can be accessed without external tuning).\cite{Matsumoto11}  These factors motivated us to search for and study the low-temperature properties of other Yb-based systems that are situated near magnetic QCPs.

Compounds with the non-centrosymmetric Zr$_2$Fe$_{12}$P$_7$-type hexagonal crystal structure (space group $P\bar{6}$) with the general formula $Ln_2T_{12}Pn_7$ (where $Ln$ = lanthanide, $T$ = transition metal, and $Pn$ = P or As) have been studied for the past 50 years.\cite{Ganglberger68,Jeitschko78,Jeitschko93,Hellmann01}  Most of these studies have involved efforts to synthesize the materials and study their crystal structure, so that their physical properties are still mostly unreported.  Recently, some interesting correlated electron behavior has been reported for several compounds with this structure, such as heavy-fermion behaviour and ferromagnetism in Sm$_2$Fe$_{12}$P$_7$,\cite{Janoscheck11} and local-moment antiferromagnetism in U$_2$Fe$_{12}$P$_7$.\cite{Baumbach11}  The temperature-magnetic field phase diagram in Yb$_2$Fe$_{12}$P$_7$, which exhibits a crossover from one NFL regime at low magnetic field to a distinct NFL regime at higher field, is particularly notable.\cite{Baumbach10}  This unconventional behavior in the Yb$_2$Fe$_{12}$P$_7$ compound may stem from spin-chain physics.\cite{in_prep}

The physical properties of Yb$_2$Ni$_{12}$P$_7$ were first studied by Cho \textit{et al.} using a polycrystalline sample that included about 8\% of Yb$_2$O$_3$ impurities.\cite{Cho98}  The main result of that study was to identify an intermediate valence (IV) state for Yb in Yb$_2$Ni$_{12}$P$_7$, where the valence of Yb was estimated to be about 2.79 based on analysis of magnetic susceptibility data.\cite{Cho98}  Recently, electrical resistivity, $\rho$, specific heat, $C$, and magnetic susceptibility, $\chi$, measurements were reported for single crystals of Yb$_2$Ni$_{12}$P$_7$.\cite{Nakano12}  These results were largely consistent with those from the polycrystalline sample, and further suggested that Yb$_2$Ni$_{12}$P$_7$ exhibits correlated electron behavior with a probable IV Yb state.\cite{Nakano12}  A study of Yb$_2$Ni$_{12}$P$_7$ under applied pressure was also recently conducted by Nakano \textit{et al.}; their results suggest the existence of a QCP and demonstrate that the Fermi-liquid crossover temperature, $T_{FL}$, decreases while the residual electrical resistivity, $\rho_0$, and the coefficient of the quadratic temperature-dependent term of $\rho(T)$, $A$, both increase with increasing pressure.

Though it has been suggested that Yb$_2$Ni$_{12}$P$_7$ may be near a QCP, there are no comprehensive studies of this compound that have been made to very low temperature; such a study is necessary to elucidate details of the possible QCP.  Motivated by these previous studies, and in order to study the low-temperature behavior of Yb$_2$Ni$_{12}$P$_7$ further, we performed measurements of $\rho$, $\chi$, and $C$ on high-quality single crystalline samples at temperatures down to 100 mK for $\rho(T)$, and 2 K for $\chi(T)$ and $C(T)$.  We also performed measurements of the thermoelectric power, $S(T)$.  In agreement with previous reports, an IV Yb state is clearly observed in the physical properties and the valence is estimated to be $\sim$ 2.76 based on analysis of the $\chi(T)$ data.  A robust quadratic temperature dependence for $\rho$ (\textit{i.e.}, $\rho \sim T^n$ where $n = 2$) is observed below $\sim$4 K, which indicates that Yb$_2$Ni$_{12}$P$_7$ has a FL ground state.  The electronic contribution to the specific heat exhibits $C_e(T)/T \sim -\ln(T)$ behavior for 5 K $< T <$ 15 K, which is consistent with NFL behavior.  Below $\sim$5 K, the upturn in $C_e(T)/T$ begins to saturate, suggesting that the system crosses over into a FL ground state.  The possibility of a crossover between FL and NFL states near $\sim$5 K is strengthened by analysis of the temperature dependence of the power-law exponent, $n(T)$, of the electrical resistivity, which becomes sub-quadratic at temperatures above 4 K.  A crossover between FL and NFL behavior is consistent with Yb$_2$Ni$_{12}$P$_7$ being in proximity to an unidentified QCP.

\section{EXPERIMENT}
\label{sec:experiment}

\indent Single-crystalline samples of Yb$_2$Ni$_{12}$P$_7$ and La$_2$Ni$_{12}$P$_7$ were prepared from elemental Yb (small dendritic pieces) or La (small chunks), Ni (small dendritic pieces), and P (small lumps).  The crystals were grown by reacting elements  with purities of 99.9\% (or better) with the initial atomic ratios Yb:Ni:P:Sn = 1:4:2:30 in a molten Sn flux.  The starting materials were sealed under vacuum in a quartz tube, heated to 1150 $^{\circ}$C, dwelled for 24 hours, and cooled slowly to 600 $^{\circ}$C over a span of 168 hours.  The Sn flux was spun off using a centrifuge.  Any residual Sn on the surface of the crystals was etched away in dilute HCl. Needle-shaped single crystals were obtained with typical dimensions of 4 $\times$ 0.5 $\times$ 0.5 mm$^3$.

Polycrystalline samples of Yb$_2$Ni$_{12}$P$_7$ for thermoelectric power measurements were prepared from elemental Yb (small dendrite pieces), Ni (powder), and P (small lumps).  The starting materials were sealed under vacuum in quartz tubes, slowly heated to 1000 $^{\circ}$C, and held at this temperature for 3 days.  After the initial reaction, the sample was ground into a powder, pressed into a pellet, and annealed at 1135 $^{\circ}$C for 3 days.  Finally, the sample was re-ground again and fired at 1135 $^{\circ}$C for 3 days.

X-ray diffraction measurements were performed on powdered single crystals using a Bruker D8 Discoverer x-ray diffractometer.  The resulting powder diffraction patterns were refined by means of Rietveld analysis\cite{Rietveld69} implemented in the program suite GSAS.\cite{Larson00}  The stoichiometry was further verified using energy dispersive x-ray spectroscopy (EDX) measurements with a FEI Quanta 600 and an INCA EDX detector from Oxford instruments.  Isothermal magnetization $M$($H$) and magnetic susceptibility $\chi$($T$) = $M$($T$)/$H$ were measured  for $T$ = 2-300 K in magnetic fields $\mu_0H$ up to 7 T using a Quantum Design (QD) Magnetic Property Measurement System.  The magnetic field, $H$, was applied both parallel and perpendicular to the crystallographic c axis.  We measured specific heat between 2 and 50 K in a QD Physical Property Measurement System DynaCool using a standard thermal relaxation technique.  Electrical resistivity measurements were carried out in a four-wire configuration in a pumped $^4$He dewar to 1.1 K and in an Oxford Kelvinox dilution refrigerator between $T$ = 100 mK and 20 K and in magnetic fields of $\mu_0H$ = 0 T and 9 T.

Measurements of the thermoelectric power were performed by applying a static temperature gradient of $\Delta T/T$ = 2\%, where $\Delta T$ was measured using two Cernox 1050 thermometers and a Lakeshore 340 Temperature Controller.  Copper leads were attached to the sample using silver epoxy in a two-wire configuration.  The thermoelectric voltage generated by the sample was measured using a Keithley 2182 Nanovoltmeter and was corrected for a background contribution arising from a slight compositional asymmetry of the alloys used in the twisted pairs of wires running from the sample to the external electronics at room temperature.

\section{RESULTS}
\label{sec:results}

\subsection{Crystal structure}

\begin{figure}
  \begin{center}
    \includegraphics[scale=0.45]{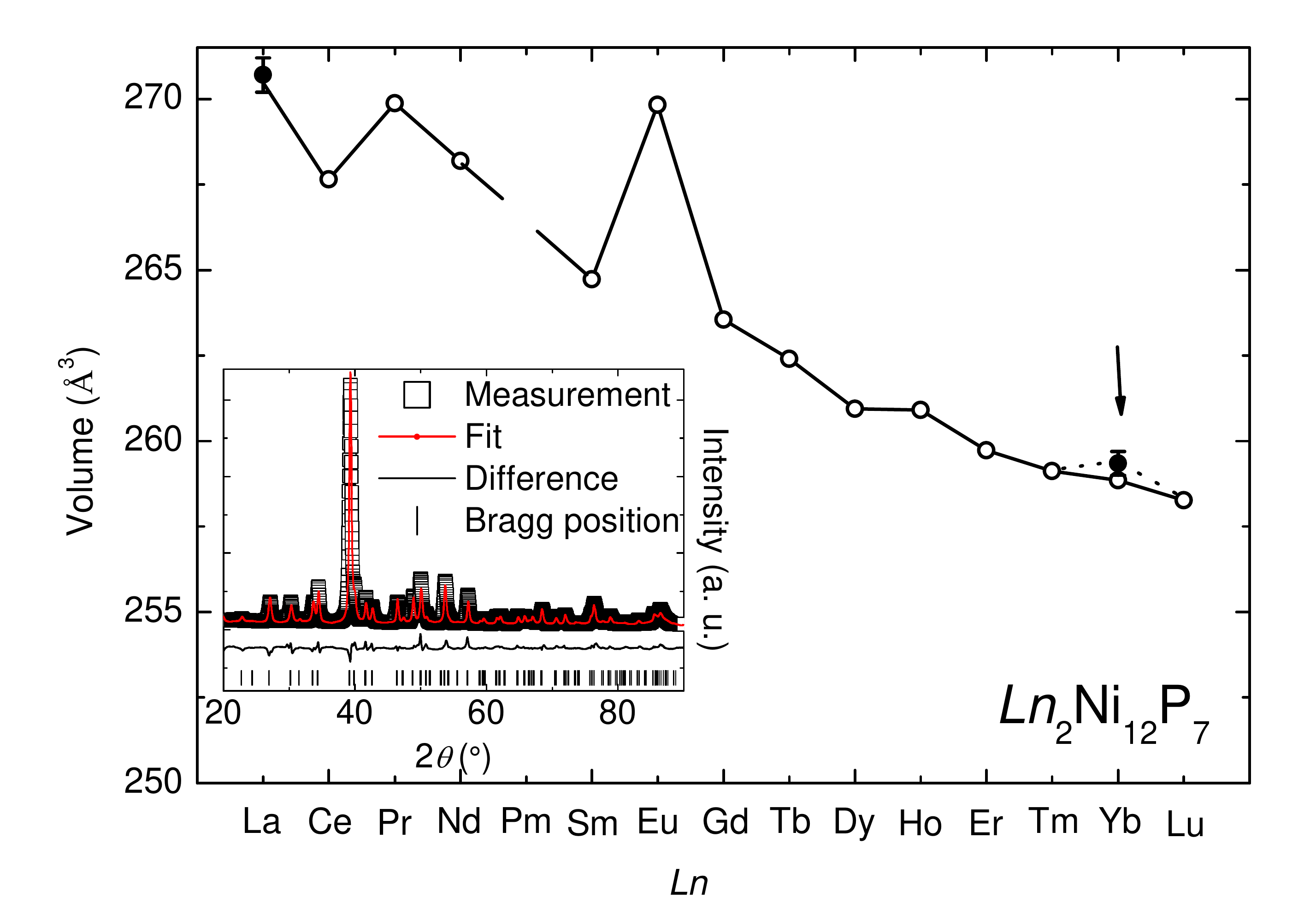}
  \end{center}
  \caption{\label{XRD} Unit-cell volume vs. $Ln$ element for $Ln_2$Ni$_{12}$P$_7$ compounds; open circles are from Ref.~\cite{Jeitschko93} and filled circles are from this work.  Inset: X-ray diffraction pattern of Yb$_2$Ni$_{12}$P$_7$ at room temperature (open squares), fit from Rietveld refinement (red line), and the difference between measured data and the fit (line).}
  \label{fig:fig1}
\end{figure}

\indent The unit-cell volume of Yb$_2$Ni$_{12}$P$_7$ was determined from powder x-ray diffraction measurements at room temperature to be 259.4 {\AA}$^3$.  In Figure~\ref{XRD}, the unit-cell volume of $Ln_2$Ni$_{12}$P$_7$ is plotted vs. lanthanide ion \textit{Ln} where open circles are taken from Ref.~\cite{Jeitschko93}.  Starting with La, one observes the typical reduction of the volume due to the contraction of the ionic radii of the lanthanides with increasing atomic number.  However, the cases of $Ln$ = Ce and Eu exhibit behavior that reflects their tendency toward assuming tetravalent and divalent states, respectively.  Our result for Yb$_2$Ni$_{12}$P$_7$ is in good agreement with other reported values within experimental uncertainty.\cite{Cho98,Jeitchko81}  According to the trend of the unit-cell volumes, the Yb ion in Yb$_2$Ni$_{12}$P$_7$ appears to be nearly trivalent at room temperature.

\subsection{Magnetic susceptibility}

\begin{figure}
  \begin{center}
    \includegraphics[scale=0.45]{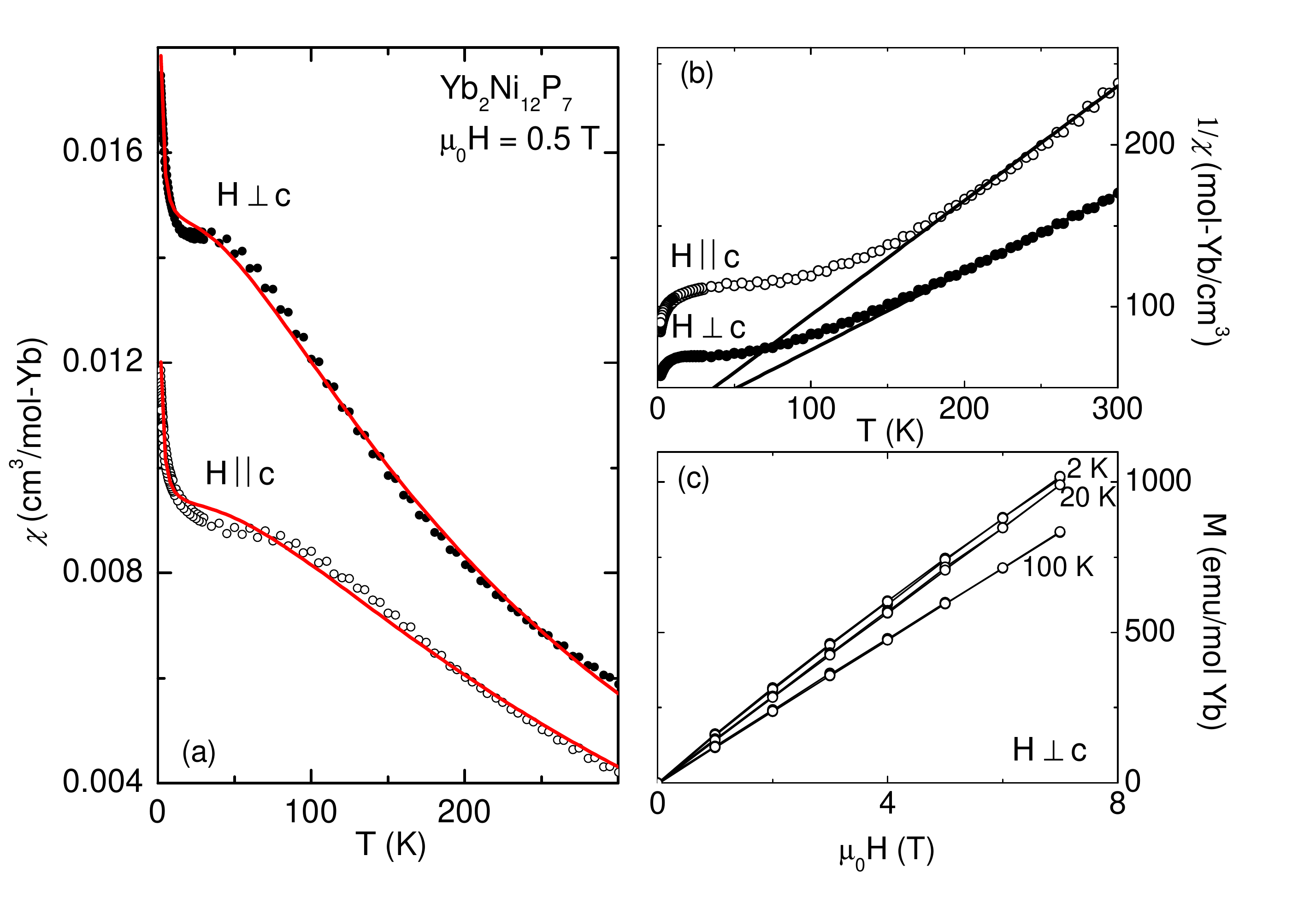}
  \end{center}
  \caption{\label{magnetic} (a) Magnetic susceptibility, $\chi$ = $M/H$, vs. temperature, $T$, for magnetic field of $\mu_0H$ = 0.5 T applied parallel ($\parallel$) and perpendicular ($\bot$) to the $c$ axis.  The $\chi(T)$ data were fitted with Eq. (1) from the ICF model as described in the text, and the best-fit results are represented by the solid lines.  (b) Inverse magnetic susceptibility, $\chi^{-1}$ = $H$/$M$, vs. $T$ for $H\parallel c$ and $H$ $\bot$ $c$.  The solid lines represent Curie-Weiss fits to the data.  (c) $M$ vs. $H$ for $H$ $\bot$ $c$ at $T$ = 2 K, 20 K, and 100 K.}
  \label{fig:fig2}
\end{figure}

\indent Magnetic susceptibility, $\chi(T)$, measurements were performed for Yb$_2$Ni$_{12}$P$_7$ in a magnetic field of $\mu_0H$ = 0.5 T with the field applied parallel and perpendicular to the c axis.  The data are shown in Figure~\ref{magnetic}.  Curie-Weiss (CW) behavior was observed at temperatures between 200 and 300 K with CW temperatures $\theta_{CW}$ = -34 K and -52 K for $H \bot c$ and $H\parallel c$, respectively (see lines in Figure~\ref{magnetic}(b)).  Effective magnetic moments, $\mu_{eff}$ = 4.1 $\mu_{B}$ and 3.4 $\mu_{B}$, for $H \bot c$ and $H\parallel c$, respectively, were also obtained from our analysis.  These two values are lower than the previously reported value of $\mu_{eff}$ = 5.1 $\mu_{B}$, which was obtained from a fit over the temperature range 50 K to 300 K; however, we note that the orientation of $H$ with respect to the c axis was not specified in that report.\cite{Nakano12}  These effective magnetic moments are intermediate between the Hund's rule values for Yb$^{3+}$ ($\mu$ = 4.5 $\mu_{B}$ for the free ion) and nonmagnetic Yb$^{2+}$ ($\mu$ = 0 $\mu_{B}$) configurations.  This result suggests that the Yb ions in this compound assume an intermediate valence.

Below 100 K, $\chi$($T$) saturates and exhibits temperature-independent Pauli paramagnetic behavior with $\chi_{0}$ = 0.014 cm$^{3}$/mol-Yb and 0.009 cm$^{3}$/mol-Yb for $H \bot c$ and $H\parallel c$, respectively.  These values are comparable to a previously reported value of $\chi_{0}$ = 0.012 cm$^{3}$/mol-Yb and suggest a moderately-enhanced effective mass for the conduction electrons.\cite{Nakano12}  Measurements of $M$ vs. $H$ are displayed in Figure~\ref{magnetic}(c), which were performed at 2 K, 20 K, and 100 K.  These temperatures are at or lower than the crossover from Curie-Weiss to Pauli paramagnetic behavior, and the $M$ vs. $H$ data are consistent with that interpretation.  This type of behavior is common for mixed-valent Ce- and Yb-based compounds as seen, for instance, in CeRhIn and YbCuAl.\cite{Adroja89,Mattens77}  The sharp rise of $\chi(T)$ below 20 K is probably due to the presence of Yb$_{2}$O$_{3}$ as a secondary impurity phase in the samples; though it could also be consistent with NFL behaviour.\cite{Cho98,Havinga73,Adroja90,Fisher00}

The behavior of $\chi$($T$) can be interpreted within the framework of the inter-configuration fluctuation (ICF) model.\cite{Sereni91,Maple74}  In the ICF model, which was applied to Yb$_2$Ni$_{12}$P$_7$ in Ref.~\cite{Cho98}, expected values for measured physical quantities (\textit{e.g.}, Yb valence, M\"{o}ssbauer isomer shift, magnetic susceptibility, etc.) are calculated by averaging over the occupied 4$f^{n}$ and 4$f^{n-1}$ multiplet states.  The energy difference between the ground state and excited-state configurations is defined as $E_{exc}$.  The  $\chi(T)$ data, and the fit to $\chi(T)$ using the ICF model, are plotted as a function of temperature for both orientations ($H\parallel c$ and $H \bot c$) in Figure~\ref{fig:fig2}(a).  The ground state of Yb (Yb$^{2+}$, 4$f^{14}$) has total angular momentum $J$ = 0 and effective magnetic moment $\mu_{eff}^{2+} = 0$, while the excited state (Yb$^{3+}$, 4$f^{13}$) has $J$ = 7/2 and $\mu_{eff}^{3+}$ = 4.54 $\mu_{B}$.  The temperature dependence of the susceptibility of the mixed-valence Yb ions in the ICF model is given by\cite{Sales75}
\begin{equation}
\chi(T) = (1-n_{imp}) \frac{N(\mu^{3+}_{\rm{eff}})^{2}P(T)}{3k_B(T + T_{vf})} + n_{imp} \frac{C}{T} + \chi_{0},
\end{equation}
where
\begin{equation}
P(T) = \frac{8}{8+\exp [E_{exc}/k_B(T + T_{vf})]}.
\end{equation}
\indent $P(T)$ is the fractional occupation probability for the 4$f^{13}$ contribution, $T_{vf}$ is the valence-fluctuation temperature, $\chi_{0}$ is a temperature-independent contribution to the magnetic susceptibility (but is not the enhanced Pauli susceptibility), and $N$ is the number of Yb ions.  The Curie term in Eq.~(1), weighted by a scale factor $n_{imp}$, is introduced to account for the upturn in $\chi(T)$ at low temperatures that is likely caused by paramagnetic impurities containing Yb$^{3+}$ ions.  The other 1-$n_{imp}$ fraction of Yb ions in the single crystals produce the intrinsic magnetic susceptibility of Yb$_2$Ni$_{12}$P$_7$ (characterized in Eq.~(1) by the ICF model term).  A least-squares fitting procedure was used to fit Eq.~(1) to the experimentally observed $\chi(T)$ data plotted in Figure~\ref{magnetic}(a).  The best fits are shown as solid lines in Figure~\ref{magnetic}(a) with the following parameters: $E_{exc}$ = 325 K, $T_{vf}$ = 140 K, $n_{imp}$ = 0.016, and $\chi_{0}$ = 5.46$\times$ 10$^{-3}$ cm$^{3}$/mol for $H$ $\bot$ $c$, and $E_{exc}$ = 453 K, $T_{vf}$ = 198 K, $n_{imp}$ = 0.013, and $\chi_{0}$ = 5.27$\times$ 10$^{-3}$cm$^3$/mol for $H\parallel c$.  The valence of the Yb ion in Yb$_2$Ni$_{12}$P$_7$ is calculated at room temperature using Eq.~(2) to be 2.79 and 2.76 for $H$ $\bot$ $c$ and $H\parallel c$, respectively.  These parameters agree reasonably well with values of $E_{exc}$ = 278 K, $T_{vf}$ = 68 K, and a valence of 2.79 that were obtained in a previous ICF model analysis of $\chi(T)$ data for a polycrystalline sample of Yb$_2$Ni$_{12}$P$_7$.\cite{Cho98}  The values of $T_{vf}$ are also comparable to values reported for other Yb-based mixed-valence compounds, such as YbAl$_{3}$.\cite{Beal80}  The best-fit $E_{exc}$ values are comparable to those of certain Yb compounds, such as YbB$_{4}$,\cite{Sales75} but are about a factor of two larger than the value for the compound YbAl$_{3}$.\cite{Beal80}

From the analysis of $\chi(T)$ data using the ICF model, we estimate an impurity ratio of $\sim 1 \%$. This value is comparable to the impurity concentration that is estimated from the entropy associated with AFM order in Yb$_2$O$_3$ in $C(T)/T$ data as discussed below. Since the contribution to $\chi(T)$ from the impurity phase can be large at low temperature, even though the impurity concentration is low, the upturn probably originates from the presence of Yb$_{2}$O$_{3}$ in the sample.  Another possible explanation for the upturn in $\chi(T)$ between 2-10 K is that it could be a manifestation of NFL behavior; this explanation would suggest that the upturn is an intrinsic property of Yb$_2$Ni$_{12}$P$_7$ rather than being extrinsic (\textit{i.e.}, from impurities).  We find that the upturn can be described by a weak power law or with a logarithmic temperature dependence of the form $\chi_{ab}(T)$ = $a$ - $b$ln($T$ - $c$) where best-fit values were found to be $a$ = 0.016 cm$^3$/mol, $b$ = 7.82$\times$ 10$^{-4}$ cm$^3$/mol, and $c$ = 1.94 K.  Such a temperature dependence for the magnetic susceptibility is a significant departure from FL behaviour; however, we cannot ignore the compelling evidence for a small amount of magnetic impurities in our samples.  While these impurities likely play a significant role in producing the low-temperature upturn, we also cannot rule out the possibility that it arises from a combination of NFL behaviour (intrinsic) and Curie-law behaviour (extrinsic).

%Other possibility to explaining the upturn behavior, the temperature range between 2 K to 10 K, could be related to NFL behavior which is the intrinsic properties of Yb$_2$Ni$_{12}$P$_7$, we find that $\chi_{\\ab}$ can be described using either power law or logarithimic functions of the forms $\chi_{\\ab}$ = $a$ - $b$ln$(T - c)$, where $a$ = 0.016 cm$^3$/mol, $b$ = 7.82*10$^{-4}$ cm$^3$/mol, and $c$ = 2.97 K. For $\chi_{\\c}$, $a$ = 0.011 cm$^3$/mol, $b$ = 7.23*10$^{-4}$ cm$^3$/mol, and $c$ = 1.94 K. From this analysis, this $T$ dependence represents a departure from FL behavior. We note that the contribution of impurity can be large at low temperature, it is possible it shows combining behavior between magnetic contribution form impurity contribution and NFL at low temperature.

%According to the analysis of ICF model we can assume the impurity ratio is between 1.31 and 1.61 percentages, which values are comparable with impurity ratio from entropy calculation. Since the impurity single can be stronger ar low temperature in this IV system, even the impurity ratio is low, the upturn behavior might be originated from this impurity phase. On the other aspect, the upturn behavior might be related with NFL behavior. However, in the view of the ICF model is fitted very well with our data and the other measurement, such as $\rho(T)$ and $C/T(T)$, shows some evidences for FL behavior in this temperature range, this upturn behavior is more likely affected on impurity in IV system.

\begin{figure}
  \begin{center}
    \includegraphics[scale=0.45]{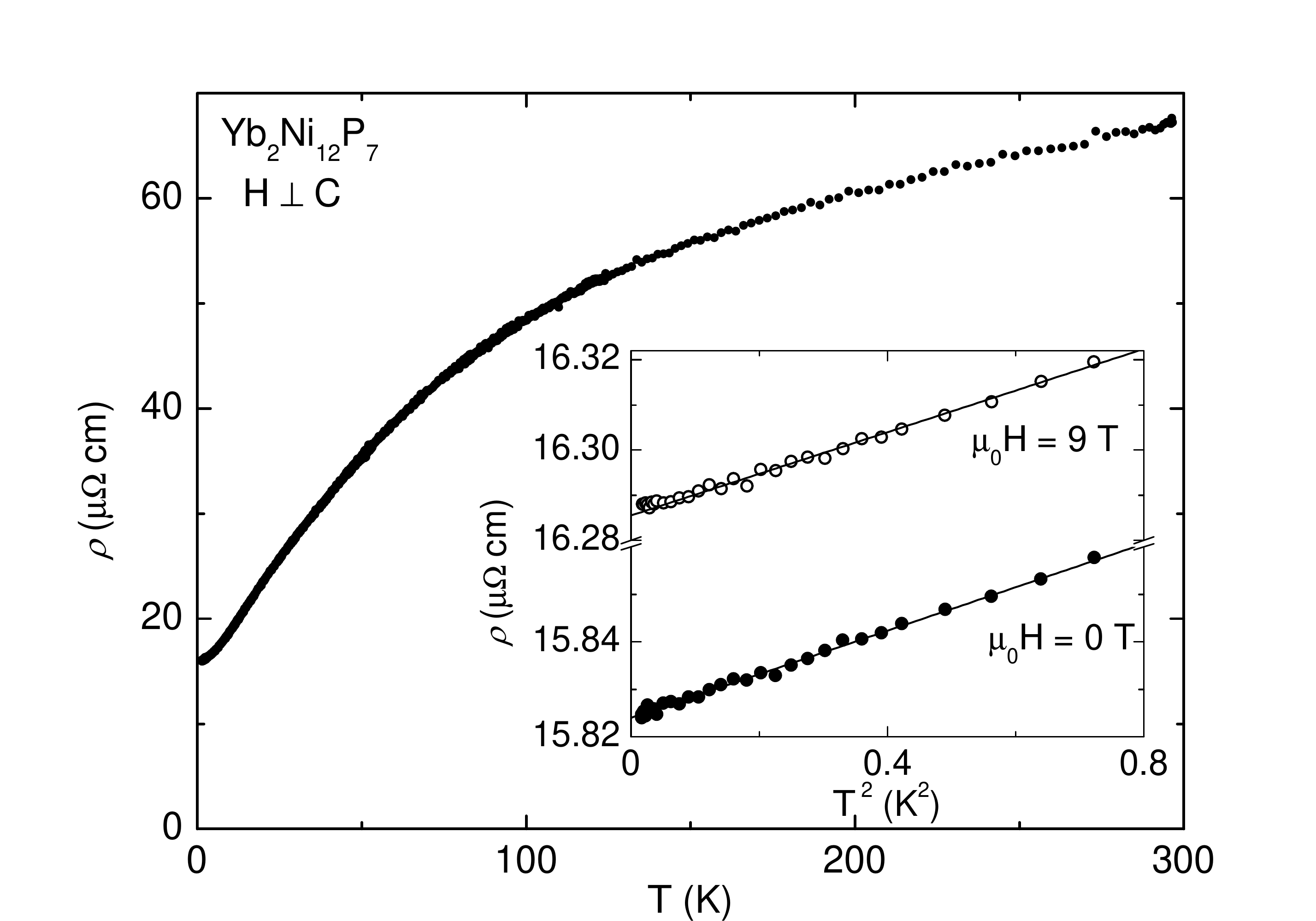}
  \end{center}
  \caption{\label{resistivity} Electrical resistivity, $\rho$, vs. temperature, $T$, for \ynp.  The inset displays $\rho$ vs. $T^2$ data for magnetic fields of $\mu_0H$ = 0 T and 9 T applied parallel to the $c$ axis.  The solid lines are fits to the data using Eq.~(3).}
  \label{fig:fig3}
\end{figure}

\subsection{Electrical resistivity}

\indent The electrical resistivity, $\rho(T)$, of Yb$_2$Ni$_{12}$P$_7$, measured in magnetic fields $\mu_0H$ = 0 T and 9 T, is shown in Figure~\ref{resistivity} between 100 mK and 300 K.  $\rho(T)$ decreases linearly between 180 K and 300 K with decreasing temperature.  At lower temperatures, there is a gentle roll off with decreasing temperature that could be related to splitting of the $J$ = 7/2 multiplet of Yb by the crystalline electric field.  The curvature in $\rho(T)$ has also been previously attributed to typical behavior for IV Yb-based compounds with low values of $E_{exc}$ and $T_{vf}$.\cite{Cho98}  To characterize $\rho$($T$, $H$) at low temperature, the data were fitted with a power law,
\begin{equation}
\rho(T) = \rho_{0} + AT^{n},
\end{equation}
where the best fit was determined from a plot of $\ln(\rho - \rho_{0})$ versus $\ln(T)$.  We note that $A$ is an important quantity in FL systems (\textit{i.e.}, for $n = 2$), but in other cases where $n \ne 2$, $A$ would just be interpreted as a coefficient.  The value of $\rho_0$ was selected to maximize the linear region of the fit extending from low $T$.  Such fits always resulted in $n$ = 2 as is emphasized in the plot of $\rho$ vs. $T^2$ shown in the inset of Figure~\ref{resistivity}, where it is apparent that the electrical resistivity follows a FL-like $T^2$ temperature dependence at low $T$.\cite{Buffat86,Andres75}  A quadratic temperature dependence for $\rho$ at low $T$ is observed in many other Yb- and Ce-based mixed-valence and heavy-fermion compounds (\textit{e.g.}, CeInPt$_{4}$ and YbAgCu$_{4}$).\cite{Malik89,Rossel87}  From these fits, we find that $\rho_0 \sim$ 15.8 $\mu\Omega$ cm and 16.3 $\mu\Omega$ cm and that $A$ = 0.0458 $\mu\Omega$ cm/K$^{2}$ and 0.0462 $\mu\Omega$ cm/K$^2$ for $\mu_0H$ = 0 T and 9 T, respectively.  These results suggest that applied magnetic fields less than $\mu_0H$ = 9 T have a negligible effect on the FL ground state.

\subsection{Specific heat}

\begin{figure}
  \begin{center}
    \includegraphics[scale=0.6]{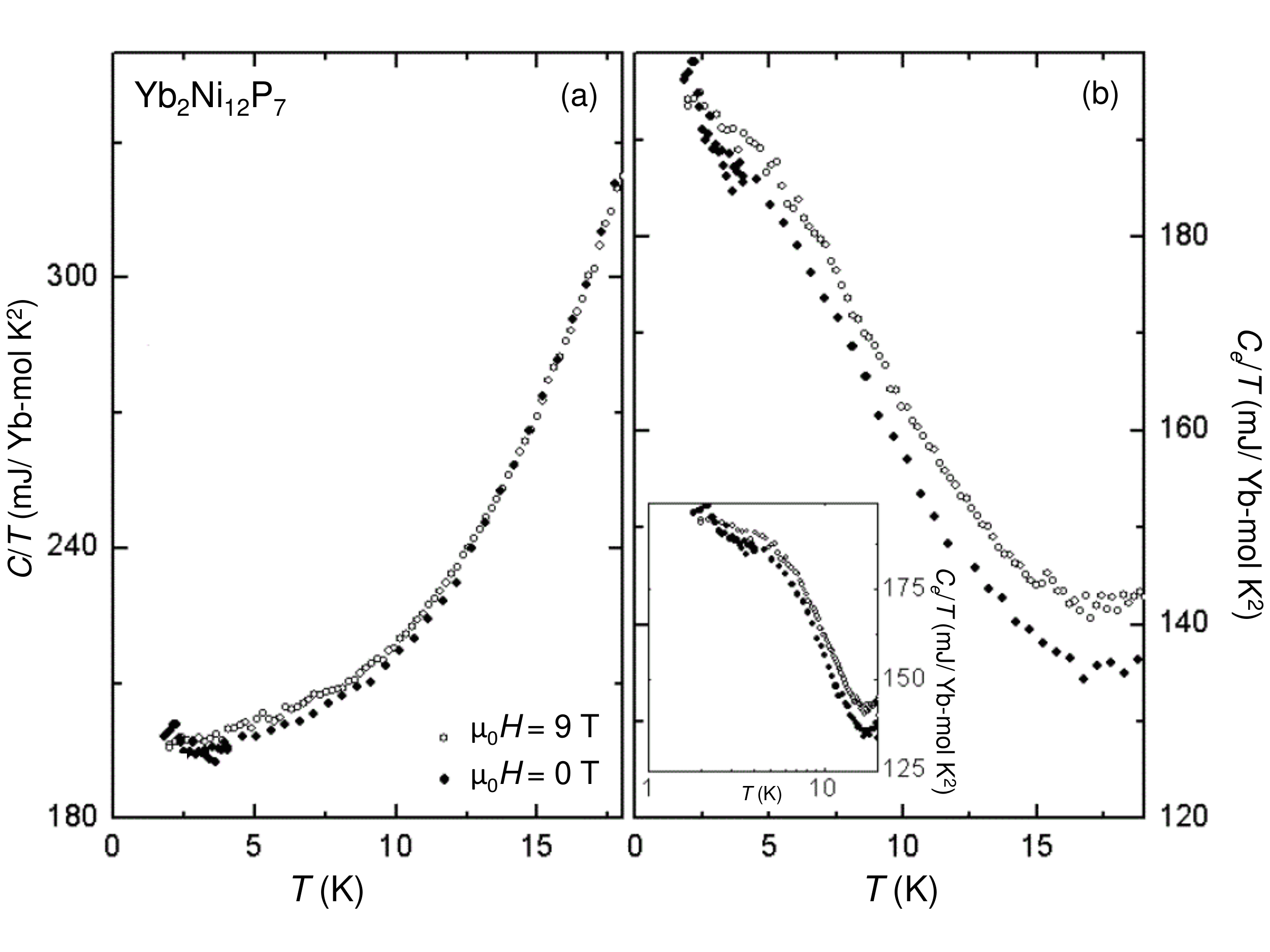}
  \end{center}
  \caption{\label{specific}(a) Specific heat, $C$, divided by temperature, $T$, for Yb$_2$Ni$_{12}$P$_7$, measured in applied magnetic fields $\mu_0H$ = 0 T and 9 T.  Filled and open circles represent $\mu_0H$ = 0 T and $\mu_0H$ = 9 T data, respectively.  (b) Electronic contribution to specific heat, $C_e/T$, obtained by subtracting the phonon contribution.  The inset displays $C_e/T$ plotted on a logarithmic temperature scale.  The upturn in $C_e/T$ has a logarithmic character down to $\sim$ 5 K, below which, it tends to saturate with decreasing $T$.}
  \label{fig:fig4}
\end{figure}

\indent The specific heat of Yb$_2$Ni$_{12}$P$_7$ is plotted in Figure~\ref{specific}(a) as $C/T$ vs. $T$ as measured in $\mu_0H$ = 0 T and 9 T applied magnetic fields.  There appears to be very little difference between the data measured in $\mu_0H$ = 0 T and 9 T magnetic fields.  The upturn in $C/T$ at low temperature prevents us from extracting the Sommerfeld coefficient, $\gamma$, from a fit of $C/T$ vs. $T^2$ at temperatures $T \le \Theta_D/50$.  If we estimate $\gamma$ by extrapolating $C/T$ to zero temperature, denoted herein as $\gamma_{0}$, we obtain $\gamma_{0} \sim$ 192 mJ/Yb mol-K$^{2}$ and 196 mJ/Yb mol-K$^{2}$ for $\mu_0H$ = 0 T and 9 T, respectively.  These values for $\gamma_0$ are comparable to other values that were reported previously.\cite{Cho98,Nakano12}  A small peak near $T_N$ = 2.3 K is observed in Figure~\ref{specific}(a) (data measured in $\mu_0H$ = 0 T) which is due to antiferromagnetic (AFM) ordering of Yb$_{2}$O$_{3}$ impurities.\cite{bauer03}  These impurities are probably present on the surface of the single crystals.  The presence of Yb$_2$O$_3$ impurities was also reported in other studies of Yb$_2$Ni$_{12}$P$_7$.\cite{Cho98}  The entropy associated with this anomaly, presumably proportional to the concentration of the impurity phase that produces it, is about 1.2$\%$ of $R\ln2$.  Figure~\ref{specific}(b) shows the electronic contribution to specific heat, $C_e/T$, obtained by subtracting the lattice contribution from $C/T$.  In the temperature range from $T$ = 2 K to 17 K, $C_e/T$ appears to increase with decreasing temperature.  This same behavior was reported in Ref.~\cite{Cho98} despite the presence of a larger Yb$_{2}$O$_{3}$ contribution (characterized by feature at $T_N \sim$ 2.3 K) than we observed in our single crystal samples.  In the inset of Figure~\ref{specific}(b), $C_e/T$ is plotted on a semi-log scale to emphasize the logarithmic character of $C_e/T \sim -\ln T$ down to $\sim$5 K.  This behavior is consistent with typical NFL behavior observed in many other systems.\cite{Seaman91,Andraka91,Gajewski94,Andraka93,Amitsuka93}  Below 5 K, the logarithmic divergence in $C_e/T$ begins to saturate which might indicate a crossover into a FL state at low temperature.  A similar situation is observed in the compound YbCo$_2$Zn$_{20}$ at $T \sim$ 0.2 K.\cite{Yamanaka11}

%Figure~\ref{fig:fig4}(b) reveals moderately-large electronic specific heat coefficients $\gamma \sim$ 137 mJ/mol-K$^{2}$ and 143 mJ/mol-K$^{2}$ for measurements in $H$ = 0 T and 9 T, respectively.  These values were obtained from a fit to the data in the temperature range $15 \leq T \leq 25$ K of the expression $C(T) = \gamma T + \beta T^3$ (see lines in Figure~\ref{fig:fig4}(b)).  The cubic term characterizes the phonon contribution from the Debye model.  We estimate the Debye temperature $\Theta_D = (12\pi 4N_{i}R/5\beta)^{1/3}$ to be 329 K and 335 K for $H$ = 0 T and 9 T, respectively; here, $N_{i}$ = 21 is the number of ions per formula unit and $R$ is the universal gas constant.

\subsection{Thermoelectric power}

\begin{figure}
  \begin{center}
    \includegraphics[scale=0.42]{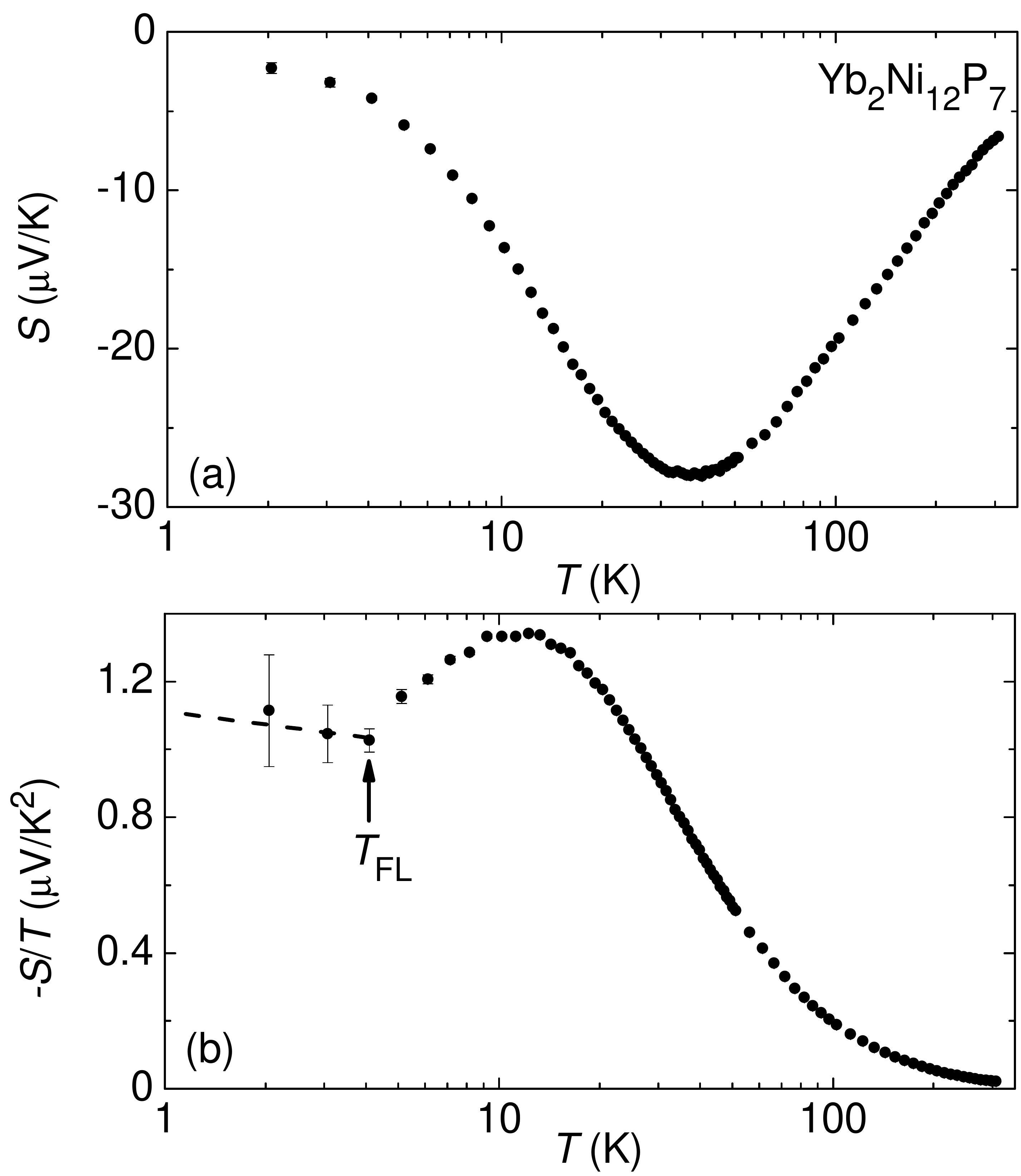}
  \end{center}
  \caption{\label{thermo}(a) Thermoelectric power, $S(T)$, for Yb$_2$Ni$_{12}$P$_7$ plotted on a semilogarithmic scale.  A large negative peak is observed in $S(T)$ near 40 K.  (b) $S(T)$ divided by temperature, $S(T)/T$, plotted on a semilogarithmic scale.  An arrow explicitly designates a temperature, $T_{FL}$, associated with a crossover from non-Fermi liquid behavior to Fermi liquid behavior for $T$ $\le$ $T_{FL}$ observed in our measurements of specific heat and electrical resistivity.}
  \label{fig:fig5}
\end{figure}

Figure~\ref{thermo}(a) displays the thermoelectric power, $S(T)$, for Yb$_2$Ni$_{12}$P$_7$ between 2 K and 300 K.  The sign of $S(T)$ is negative over the entire measured temperature range.  Large, negative values for $S(T)$ are commonly observed in other Yb-based Kondo lattice systems.\cite{Andreica99, Deppe08}  The deep minimum near $\sim$40 K could be related to spin fluctuations\cite{Andreica99, Friedemann08} like in YbCu$_2$Si$_2$.  However, a single minimum is also predicted for a generic Yb-based Kondo lattice system in which hybridization between localized and itinerant electron states is strong enough to facilitate an intermediate Yb valence.\cite{Zlatic05}  This latter scenario is consistent with the evidence for an intermediate Yb valence in Yb$_2$Ni$_{12}$P$_7$ from our magnetic susceptibility data.

Figure~\ref{thermo}(b) displays $S(T)$ data divided by $T$, -$S(T)/T$, plotted on a logarithmic $T$ scale.  The -$S(T)/T$ data increase upon cooling for $T \ge$ 13 K, reaching a maximum value of 1.34 $\mu$V/K$^2$ at $T_{max} \sim$ 13 K.  The maximum value of $S(T)/T$ is almost 2 orders of magnitude larger than the maximum values for simple metals such as Cu where $S/T$ $\sim$ -30 nV/K$^2$.\cite{Behnia04}  Enhanced values of $S(T)/T$ have been reported for many HF compounds\cite{Izawa07, Hartmann10, Behnia04} and are considered to be closely related to the strongly-enhanced values of $C(T)/T$. As can be seen in Figure~\ref{thermo}(b), -$S(T)/T$ exhibits a break in slope at a temperature we denote $T_{FL}$.  These observations are qualitatively consistent with a FL ground state as previously inferred from the electronic contribution to specific heat, $C_e/T$, in which a potential crossover between NFL and FL behavior is observed at $T_{FL}$.  Though there is reasonable evidence for such a crossover in both $C_e/T$ and $\rho(T)$ data (see next section), we note that we are unable to definitively identify any evidence for NFL behavior in our $S/T$ data for $T > T_{FL}$.

If we extrapolate $S(T)/T$ to zero temperature, we obtain a value of $-S_0/T$ $\sim$ 1.1(2) $\mu$V/K$^2$.  A FL state can be characterized by the ratio of $S_0(T)/T$ to $\gamma$.\cite{Behnia04, Grenzebach06, Zlatic07}  A ``quasi-universal'' ratio, $q = (N_{A}e/\gamma)(S_0/T) \approx \pm$1, is expected to be obeyed for FL systems where $N_A$ is Avogadro's number and $e$ is the charge of an electron.  The sign of $q$ depends on the dominant type of charge carriers.  Although a single band and scattering process is generally insufficient to explain the strong correlation effects in materials like HF systems, given that $C(T)/T$ and $S(T)/T$ are most sensitive to the position of the heavy band, a quasi-universal ratio is expected to hold at low temperature.\cite{Miyake05, Kontani03}  Using our value of $S_0$($T$)/$T$ $\sim$ -1.1(2) $\mu$V/K$^2$ and $\gamma_0 \sim$ 192 mJ/Yb mol-K$^{2}$, we obtain $q$ $\sim$ -0.6(1).  This calculated value of $q$ is close to the expected value for an Yb-based FL system, supporting the evidence from specific heat for a FL ground state in Yb$_2$Ni$_{12}$P$_7$.

\section{DISCUSSION}
\label{sec:discussion}

\begin{figure}
  \begin{center}
    \includegraphics[scale=0.45]{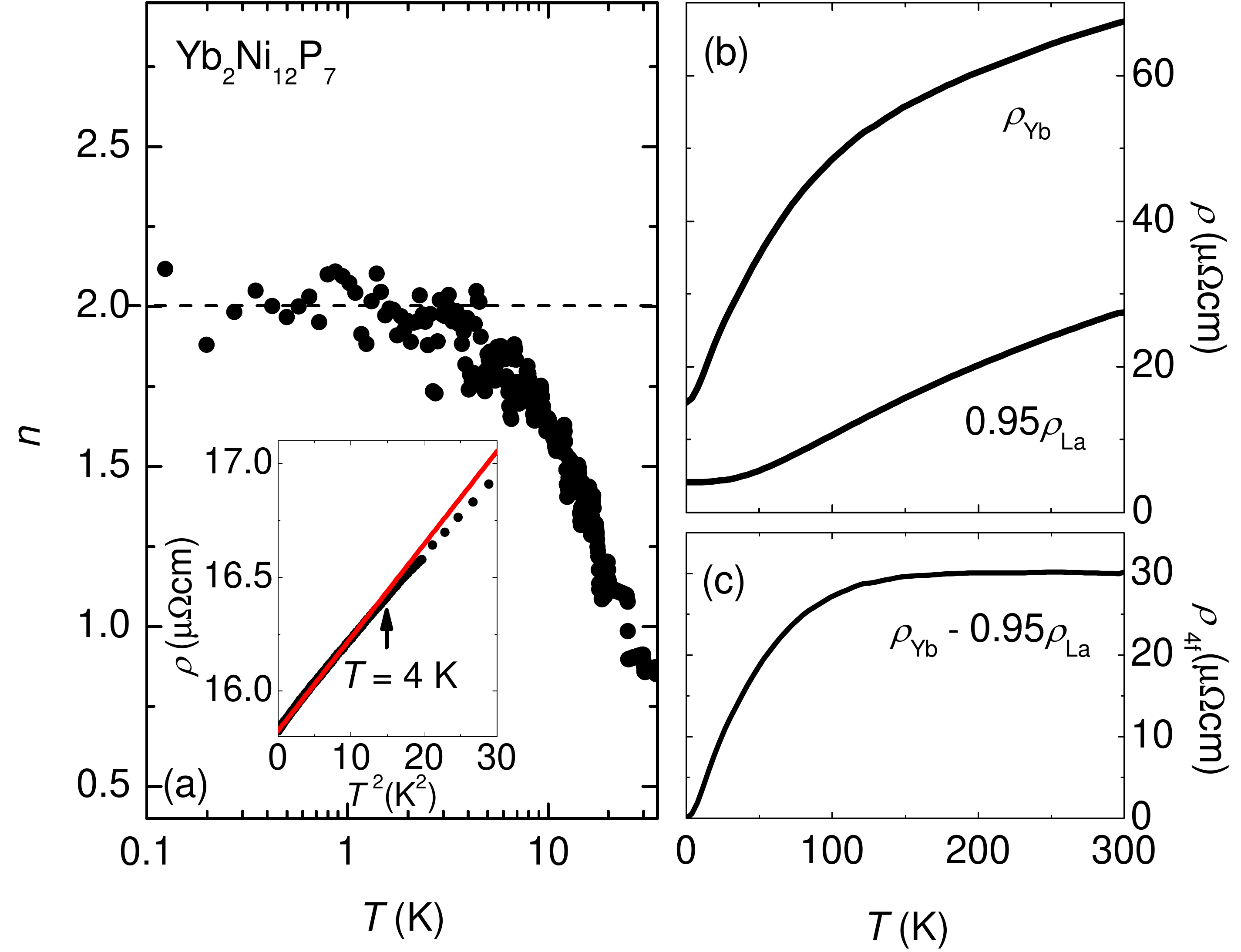}
  \end{center}
  \caption{\label{power} (a) The exponent, $n$, defined in the expression, $\rho = \rho_0 + AT^n$, and calculated as described in the text, is plotted vs. $\log T$.  Inset: $\rho$ vs. $T^2$ data; the red line emphasizes the temperature range over which $\rho$ exhibits a quadratic temperature dependence.  (b) $\rho(T)$ data for Yb$_2$Ni$_{12}$P$_7$ and La$_2$Ni$_{12}$P$_7$.  The $\rho(T)$ data for La$_2$Ni$_{12}$P$_7$ have been scaled by a factor of 0.95 so that the slope ($d\rho/dT$) at high temperature is equal to the slope of the $\rho(T)$ data for Yb$_2$Ni$_{12}$P$_7$.  (c) The $4f$ electron contribution to $\rho(T)$ in Yb$_2$Ni$_{12}$P$_7$, $\rho_{4f}$, plotted vs. $T$.  The $\rho_{4f}$ contribution was obtained by subtracting the scaled $\rho(T)$ data for La$_2$Ni$_{12}$P$_7$ from the $\rho(T)$ data for Yb$_2$Ni$_{12}$P$_7$.}
  \label{fig:fig6}
\end{figure}

\indent In order to evaluate potential HF behavior, we calculated the Kadowaki-Woods ratio, $R_{KW} = A/\gamma^2$, where $\gamma$ is the coefficient of electronic specific heat and $A$ is the coefficient of the $T^{2}$ contribution to $\rho$.  In the original treatment, Kadowaki and Woods reported a universal value $R_{KW} \sim$ 10$^{-5}$ $\mu\Omega$ cm (mol K$^{2}$ mJ$^{-1}$)$^{2}$ for all HF compounds.\cite{Kadowaki86}  More recently, it has been found that many (but not all) Yb-based HF systems are characterised by $R_{KW}$ values that can be two orders of magnitude smaller than the conventional Kadowaki-Woods ratio value.  This result is related to the fact that the ground-state degeneracy of $N$ = 8 for Yb modifies the conventional Kadowaki-Woods ratio so that it becomes $R^{*}_{KW} \approx 1 \times 10^{-5}/(\frac{1}{2}N(N-1))$ $\mu\Omega$ cm (mol K$^{2}$ mJ$^{-1}$)$^{2}$ = 0.36 $\times 10^{-6}$ $\mu\Omega$ cm (mol K$^{2}$ mJ$^{-1}$)$^{2}$.\cite{Tsujii05}  Using our values of $\gamma_0 \sim$ 192 mJ/Yb mol-K$^{2}$ and $A \sim$ 0.0458 $\mu\Omega$ cm/K$^{2}$, both obtained for $\mu_0H$ = 0 T, we calculate $R_{KW} \sim$ 1.24 $\times$ 10$^{-6}$ $\mu\Omega$ cm (mol K$^{2}$(mJ)$^{-1}$)$^{2}$.  This value of $R_{KW}$ suggests that Yb$_2$Ni$_{12}$P$_7$ has a heavy FL ground state.  Similar results are obtained in the case of $\mu_0H$ = 9 T where $R_{KW}$ $\sim$ 1.20 $\times$ 10$^{-6}$ $\mu\Omega$ cm (mol K$^{2}$(mJ)$^{-1}$)$^{2}$ (calculated using $\gamma_0 \sim$ 196 mJ/Yb mol-K$^{2}$ and $A \sim$ 0.0462 $\mu\Omega$ cm/K$^{2}$).  Identifying the ground state of Yb$_2$Ni$_{12}$P$_7$ as being a heavy FL seems to be at odds with the logarithmic divergence with decreasing $T$ in $C_e/T$ as plotted in the inset of Figure~\ref{fig:fig4}(b).  The $C_e/T \sim -\ln T$ behavior for $5 \le T \le 15$ is a typical characteristic of NFL behavior.\cite{maple10}  However, below $T \sim 5$ K, $C_e/T$ tends to saturate with decreasing $T$, which is indicative of a crossover between FL and NFL behavior in the vicinity of $T \sim$ 5 K.

To further investigate the possibility that there is a crossover between FL and NFL behavior in Yb$_2$Ni$_{12}$P$_7$, an analysis of the evolution of the exponent, $n$, (as defined in Eq.~(3)) with temperature was conducted and is plotted in Figure~\ref{power}(a).  It is important to note that Eq.~(3) is only valid in temperature regions where electron-phonon scattering and other contributions are negligible (\textit{i.e.}, at low temperature).  Since electron-phonon scattering contributes an additive term to $\rho(T)$ that is a power law in temperature according to the Bloch-Gr\"{u}neisen theory, it is important to subtract this contribution so that $n(T)$ can be meaningfully calculated.  Measurements of $\rho(T)$ for Yb$_2$Ni$_{12}$P$_7$ and a non-magnetic reference compound, La$_2$Ni$_{12}$P$_7$, are shown in Figure~\ref{power}(b).  The $\rho(T)$ data for La$_2$Ni$_{12}$P$_7$ have been scaled by a factor of 0.95, so that their slope matches the slope of $\rho(T)$ data for Yb$_2$Ni$_{12}$P$_7$ at high temperatures.  The result of subtracting the scaled $\rho(T)$ data of La$_2$Ni$_{12}$P$_7$ from data for Yb$_2$Ni$_{12}$P$_7$ is plotted in Figure~\ref{power}(c).  The data plotted in Figure~\ref{power}(c) are expected to obey a $\rho_{4f} \sim T^n$ temperature dependence at moderately-low temperatures.  The exponent, $n(T)$, is calculated by differentiating $\ln\rho(T)$ data with respect to $\ln T$ (\textit{i.e.}, $n = d\ln(\rho-\rho_0)/d\ln T$).  As was already shown in the inset of Figure~\ref{resistivity}, $n$ = 2 at low temperature, indicating a FL ground state; however, above $T \sim$ 4 K, $n$ continuously deviates from $n$ = 2, rapidly becoming sub-quadratic.  This is consistent with $\rho$ vs. $T^2$ data in the inset of Figure~\ref{power}(a) in which a red line emphasises the deviation from a quadratic temperature dependence above $T \sim$ 4 K.  It would be reasonable to expect the quadratic temperature dependence of $\rho_{4f}$ to extend to temperatures on the order of $T_{vf}$, which plays the role of an effective Fermi temperature in Yb$_2$Ni$_{12}$P$_7$.  Since 4 K is much smaller than the $T_{vf}$ values we extracted from our analysis of the magnetic susceptibility data (\textit{i.e.}, $T_{vf}$ = 140 K or 198 K depending on the direction of the applied magnetic field), we feel that this deviation from a quadratic temperature dependence is an important observation.  In analogy with the behavior of $C_e/T$, this behavior is consistent with a crossover between a FL ground state and NFL behavior in the vicinity of $T \sim$ 4 K.

Evidence for a crossover between a FL ground state and NFL behavior is observed in both $C_e(T)/T$ and $\rho(T)$ data.  These mutually-reinforcing results strongly suggest the possibility that Yb$_2$Ni$_{12}$P$_7$ is in close proximity to a QCP.  The possibility of a nearby QCP has also been discussed by Nakano \textit{et al.} in their study of Yb$_2$Ni$_{12}$P$_7$ under applied pressure.\cite{Nakano12}  There are several types of QCP phase diagrams that have been observed in other systems;\cite{maple10} one commonly observed scenario involves the suppression of an ordered FL state to zero temperature using a non-thermal control parameter $\delta$ (\textit{i.e.}, pressure, magnetic field, chemical composition, etc.) such that a QCP is observed at $\delta_c$.  This QCP may or may not be protected by a superconducting phase, but in either case, increasing $\delta$ further leads to a region where there is a crossover between a FL state at low temperature and NFL behavior at higher temperature.\cite{maple10}  It is possible that Yb$_2$Ni$_{12}$P$_7$ is in such a region where $\delta > \delta_c$.  Further work tuning Yb$_2$Ni$_{12}$P$_7$ with applied pressure, chemical substitution, and magnetic field will be necessary to observe and identify any QCP that may be present in this system.

The possibility of tuning a system containing IV Yb ions to a QCP or for such a system to exhibit NFL behavior is relatively uncommon.  However, we note that the Yb ions in $\beta$-YbAlB$_4$, a compound that exhibits NFL behavior and is the only known Yb-based HF superconductor,\cite{Nakatsuji08} have a strongly-intermediate Yb valence of 2.75.\cite{Okawa10}.  This is comparable to the valences of 2.76 and 2.79 that we obtained for Yb in Yb$_2$Ni$_{12}$P$_7$ from our analysis of magnetic susceptibility data.  Furthermore, evidence for a zero-magnetic field QCP,\cite{Matsumoto11} protected by the superconducting state below $T_c$ = 0.08 K, has also been observed in $\beta$-YbAlB$_4$.  Quantum valence criticality is also believed to be responsible for the NFL behavior observed in YbRh$_2$(Si$_{0.95}$Ge$_{0.05}$)$_2$ \cite{Watanabe10} and has also been proposed to be the mechanism driving NFL behavior and unconventional superconductivity in the system CeCu$_2$(Si$_{1-x}$Ge$_x$)$_2$.\cite{Yuan06}  Therefore, it appears that Yb$_2$Ni$_{12}$P$_7$ could be the newest member of a growing list of IV systems that are near a QCP or that exhibit NFL behavior.

\section{CONCLUDING REMARKS}
\label{sec:concluding remarks}
\indent From measurements of $\rho(T, H)$, $\chi(T, H)$, $C(T, H)$, and $S(T)$ on high-quality single-crystalline samples, we are able to conclude that Yb$_2$Ni$_{12}$P$_7$ exhibits a moderately-heavy FL ground state with an enhanced value for $\gamma_{0} \sim$ 192 mJ/Yb mol-K$^2$ and a large coefficient $A$ of the $T^2$ term in the electrical resistivity.  These results lead to a Kadowaki-Woods ratio value of $R_{KW} \sim$ 1.24 $\times$ 10$^{-6}$ $\mu\Omega$ cm (mol K$^{2}$(mJ)$^{-1}$)$^{2}$.  An IV Yb state in Yb$_2$Ni$_{12}$P$_7$ is inferred from an analysis of $\chi(T)$ data within the context of the ICF model, where an Yb valence of 2.76 is obtained.  An intermediate Yb valence is also suggested by the character of $S(T)$.  A logarithmic divergence of the electronic contribution to specific heat (\textit{i.e.}, $C_e/T \sim -\ln T$) above $\sim$5 K is strong evidence for NFL behaviour; however, $C_e/T$ saturates below 5 K, becoming less temperature-dependent indicating there is a FL ground state.  An analysis of the power-law exponent, $n(T)$, characterizing the electron-electron scattering contribution to the electrical resistivity yields $n$ = 2 at low temperature, which also indicates a FL ground state; however, above $T \sim$ 4 K, $n$ continuously decreases and reaches $n$ = 1 at $\sim$ 20 K.  This behavior is consistent with a crossover in the vicinity of $\sim$ 5 K from a NFL to a FL ground state, strongly suggesting the possibility that Yb$_2$Ni$_{12}$P$_7$ is in close proximity to a QCP.  Further efforts to tune Yb$_2$Ni$_{12}$P$_7$ with applied pressure, chemical substitution, and/or magnetic field will be necessary to observe and then study the potential QCP.

\ack

Sample synthesis and screening for superconductivity was funded by the Air Force Office of Scientific Research MURI (Grant No. FA9550-09-1-0603), sample characterization and physical properties measurements were supported by the US Department of Energy (Grant No. DE-FG02-04-ER46105), and low-temperature measurements were funded by the National Science Foundation (Grant No. DMR-1206553). Research at California State University, Fresno was supported by the National Science Foundation {Grant No. DMR-1104544}.

\end{document}